\documentclass[10pt,letterpaper,twocolumn]{article} 

\usepackage{ol2}
\usepackage[draft]{hyperref}
\usepackage{amsmath}
\usepackage{epsfig}
\usepackage{color}
\begin{document}

\twocolumn[ 

\title{Third Harmonic Generation in Cuprous Oxide:  Efficiency Determination}


\author{Laszlo Frazer,$^{1,*}$ Richard D. Schaller,$^{2,3}$ Kelvin Chang,$^3$ John B. Ketterson,$^{1}$ and Kenneth R. Poeppelmeier$^{3,4}$}

\address{
$^1$ Department of Physics, Northwestern University, \\ 2145 Sheridan Road, Evanston, IL 60208-3112, USA \\
$^2$ Center for Nanoscale Materials, Argonne National Laboratory, \\ 9700 South Cass Avenue, Building 440, Argonne, IL 60439, USA \\
$^3$ Department of Chemistry, Northwestern University, \\ 2145 Sheridan Road, Evanston, IL 60208-3112, USA \\
$^4$ Chemical Sciences and Engineering Division, Argonne National Laboratory, \\ 9700 South Cass Avenue, Argonne, IL 60439, USA \\
$^*$Corresponding author: ol@laszlofrazer.com
}

\begin{abstract}
	The efficiency of third harmonic generation in cuprous oxide was measured.  Intensities followed a non-cubic power law which indicates nonperturbative behavior.  Polarization anisotropy of the harmonic generation was demonstrated and related to the third order susceptibility.  
The results will influence the understanding of harmonic generation in centrosymmetric materials and are potentially relevant to device design and the interpretation of exciton behavior.
\end{abstract}


 ] 

\noindent Cuprous oxide (Cu\begin{math}_2\end{math}O) has been shown to have a relatively large \begin{math} \chi^{(3)} \end{math} for a material having a bandgap in the visible region \cite{mani2009large,mani2010nonlinear}.  Its cuprite structure is centrosymmetric which requires \begin{math}
	 \chi^{(2)}=0
 \end{math} and hence no second harmonic generation.  An unusual feature is a two-photon absorption process resulting in the production of 1s orthoexcitons \cite{sun2001production,yoshioka2006dark,fu2011third,frazer2013unexpectedly}.
Numerous unusual optical properties of cuprous oxide are based on the even parity of both the valence and the conduction bands.  Since photons have odd parity, optical transitions between the bands are not allowed.  The exceptions involve phonons and a spectrally narrow \cite{frohlich1991coherent} quadrupole transition to the 1s orthoexciton polariton.  Since third harmonic generation originates from virtual transitions across the gap \cite{moss1990band}, it is interesting to investigate this process in a material where real transitions are so heavily restricted.  


In a perturbative regime, harmonic generation efficiency follows a power law with an exponent equal to the order of the harmonic.  
\begin{align}
	I_{n\omega}&\propto I_\omega^{n}\label{law}
\end{align}
Many measurements find lower exponents \cite{dichiara2012scaling}. In simpler cases such as gasses, scaling is well described \cite{macklin1993high} by a nonperturbative model \cite{l1992calculations}.

The relevance of three photon processes to excitons in cuprous oxide was first inferred via the power dependence of stressed cuprous oxide paraexciton luminescence \cite{liu2004excitons}.  Power law behavior was observed with an exponent \begin{math}
	 2.8\pm0.2
 \end{math}.  In a Lyman transition study, results were consistent with cubic behavior\cite{ideguchi2008coherent}.  Using the Z-scan method and direct detection, an exponent of \begin{math}
	 1.8
 \end{math} was observed \cite{mani2010nonlinear}.  

 Nonperturbative behavior is typically realized in measurements of the higher harmonics because the two phenomena are caused by intense pumping.  Such investigations in crystals began with ZnO \cite{ghimire2010observation}.  While harmonic generation in cuprous oxide begins with the third harmonic, we will show that the magnitude is better described by a nonperturbative scaling.


Four earlier reports used 1,240 nm excitation \cite{liu2004excitons}, involved a possibly third-photon photoionizing 1,220 nm pump and 10,700 nm probe \cite{ideguchi2008coherent}, and operated with a 1,300 to 1,800 nm fundamental \cite{mani2009large,mani2010nonlinear}.  Our experiment is able to access the transparent region for third harmonic generation in cuprous oxide at low temperature, which includes fundamental wavelengths longer than 1,820 nanometers.

Previous results on the nonlinear optical constants include the second order index of refraction measured at 1,064 nm\cite{mani2009large} and two-photon absorption coefficient at 1,220 nm \cite{mani2010nonlinear}.  In the past,  \begin{math} \chi^{(3)} \end{math} effects in cuprous oxide were tacitly assumed to be isotropic. We will show that third harmonic generation is not. 

Six different cuprous oxide single crystals were used to study third harmonic generation behavior. Three synthetic samples were grown via the floating zone method\cite{chang2013removal} using at least 99.9\%, 99.99\%, and 99.999\% pure copper metal as starting material. Another high purity crystal was received from A. Revcolevschi\cite{schmidt1974growth}. It was grown in a floating zone furnace using 99.999\% pure starting material. Two geological samples were also used.   Geological samples were included for comparison because they are commonly studied in many laboratories.

\begin{figure}[htb]
\centerline{\includegraphics[width=7.5cm]{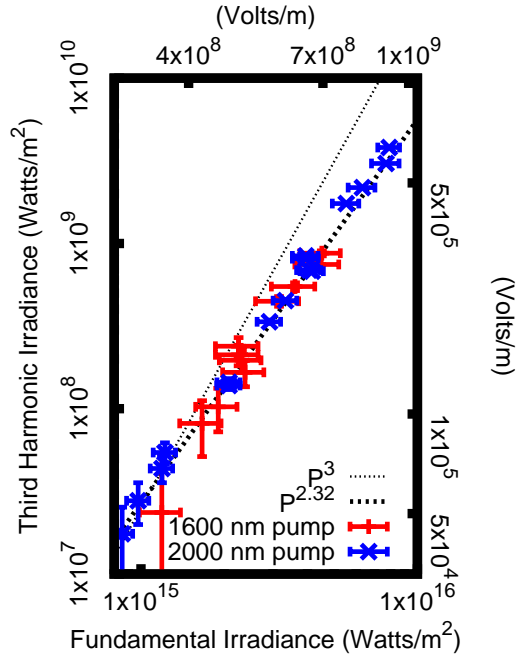}}
\caption{Dependence of third harmonic generation on fundamental laser peak irradiance.  The cubic curve shows that if a perturbative gaussian beam model with coherence length 10 $\mu$m is used, a $\chi^{(3)}$ factor about  $10^4$ less than the correct value\cite{mani2009large} would be inferred owing to non-cubic behavior at the high irradiances used here.\label{power}  The data are eleven standard errors from perturbative behavior.
}
\end{figure}

The third harmonic generation of pump wavelengths from 1,200 to 2,000 nm was investigated in multiple samples to verify reproducibility. A 
35
femtosecond OPA light source was focused on polished faces of samples in a vacuum at room temperature and the absolute magnitude of the transmitted third harmonic generation was measured with a silicon power meter. The high intensity of the pump laser allowed complete evaluation of the efficiency of third harmonic generation. {In these measurements, the fundamental is not substantially attenuated by harmonic generation.} Similar results were obtained for all samples measured for all pump wavelengths that were examined. The results from a [100] face of the crystal received from A. Revcolevschi were selected for presentation in Figure \ref{power} because its 8 mm thickness could best occupy the intense region of the beam. The beam had a 64 \begin{math}\mu\end{math}m minimum radius. 
The best power stability coefficient of variation observed was 1.5\% after harmonic generation.

\begin{figure}[htb]
\centerline{\includegraphics[width=7.5cm]{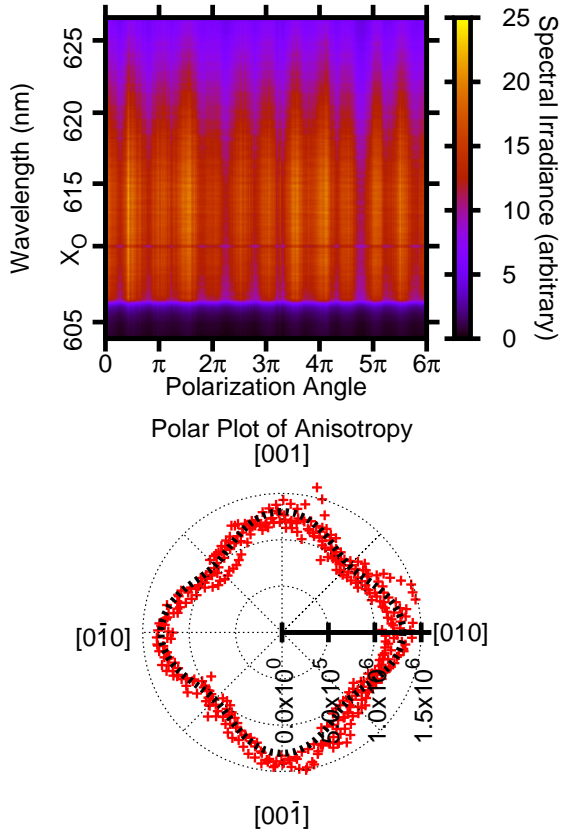}}
\caption{Dependence of third harmonic generation on fundamental laser polarization, with the electric field measured from [010].\label{polar}
The anisotropy is 22 times the standard error.  The wavelength-integrated polar plot of the data also graphs the theory shown in Expression (\ref{anisotropy}).  In the polar plot, isotropic results would form a circle.
}
\end{figure}

	The dependence of third harmonic generation on the polarization of the laser was also examined on multiple samples for reproducibility. All samples gave similar results.  Measurements from a geological sample illuminated along the [100] axis are reported in Figure \ref{polar}.  This sample gave the clearest results owing to a longer integration time.

The absolute power dependence of third harmonic generation is presented in Figure \ref{power}, which shows that large quantities of third harmonic were generated.  The overall relationship between the fundamental intensity and the third harmonic intensity can be described by a power law with an exponent lying between those reported in previous measurements \cite{liu2004excitons,mani2010nonlinear,ideguchi2008coherent}. Only a mild wavelength dependence of the third harmonic signal was observed at room temperature.  

Figure \ref{polar} shows that the third harmonic signal is in fact dependent on the polarization of the laser.  In this figure, light centered near 1,800 nm from the OPA was used to illuminate a geological cuprous oxide sample along the [100] axis. 
	The sample was cooled in vacuum to approximately 2.7 Kelvin.    The third harmonic signal spectrum was collected for various orientations of a half-wave plate located between the OPA and the sample.

	At short wavelengths, the third harmonic is greatly reduced by phonon-assisted absorption, which extends well below the bandgap energy \cite{baumeister1961optical}.  A narrow 1s quadrupole orthoexciton-polariton absorption line (\begin{math} X_O \end{math}) is also observed.  This demonstrates exciton production via the third harmonic without an intermediate electron-hole state.  Absorption studies in cuprous oxide do not show this line (except in \cite{ito1998single}, where it is weak), but it is clearly visible in our relatively thick samples.  Polarization independence of this line is restricted to the [100] axis \cite{liu2004excitons}.  Since two-photon luminescence from this line is narrow even when pumping with spectrally broad lasers \cite{shen1996second,mani2010nonlinear}, it is notable that absorption of the third harmonic photons follows the pattern of white light.  At long wavelengths, third harmonic generation displays the approximately Gaussian spectral profile of the laser.
Supercontinuum generation and plasma formation were not observed.

		For normal incidence on a [100] cubic crystal face and with \begin{math}
			\phi
		\end{math} as the angle between the electric field and the [010] axis, the total third harmonic generation is proportional to \cite[Equation 39]{sipe1987phenomenological}
\begin{align}
	&2\left( 3\chi_{1212}^{\left( 3 \right)} \right)^2+\left( A_0\frac{g_{3b}}{N} \frac{\chi_{1111}^{(3)}-3\chi_{1212}^{(3)}}{4}\right)^2 \cos^2(2\phi)
	\label{anisotropy}
\end{align}
This is compared with the data in Figure \ref{polar}.
In our experiment, no polarizer was used so that all the third harmonic generation was collected; therefore a factor of 
\begin{math}
	\sin^2(2\phi)
\end{math} found in \cite{sipe1987phenomenological} is omitted in the second term.
\begin{math} A_0, g_{3b}\end{math} and \begin{math} N \end{math} are as defined in \cite{sipe1987phenomenological}.

The measured part of third harmonic which is polarization-independent  is
\begin{align}
	\frac{
	2\left( 3\chi_{1212}^{\left( 3 \right)} \right)^2
	}{
	2\left( 3\chi_{1212}^{\left( 3 \right)} \right)^2
	+\left( A_0\frac{g_{3b}}{N} \frac{\chi_{1111}^{(3)} -3\chi_{1212}^{(3)}}{4}\right)^2
	}
	&=0.78\pm 0.01\label{prop}
\end{align}
Since in a cubic crystal the polarization is never more than \begin{math} \frac{\pi}{4} \end{math} from the axis with maximal harmonic generation,  we have a quantity \begin{math} 
	\ge
	\cos\left( \frac{\pi}{4} \right)
\end{math} in Equation (\ref{prop}).  The highest possible value, which would occur in an isotropic material, is one.

 The combined third harmonic signal (excluding linear absorption) for a fundamental measured in Watts/m\begin{math}^2\end{math} was found to be
\begin{align}
	I_{3\omega}&= (4.41\pm 0.08)\cdot 10^{-28} I_{\omega}^{2.32\pm0.06}\label{efficiency}&\\\nonumber&\times\left( \cos^2\left( 2\phi \right)+3.6\pm 0.2 \right)\text{ Watts/m}^2
\end{align}
where the overall coefficient was obtained by assuming the exponent is exact, but not vice versa. {The experimental results should be computable from the quantum mechanical polarization \cite{macklin1993high,l1992calculations} and a model of the crystal. Suitable numerical methods would be a useful area of future research.} 
The exponent is significantly nonperturbative and the anisotropy is significant.

The coherence length and beam-sample geometry impact the efficiency of harmonic generation. A coherence length of 10 \begin{math}\mu\end{math}µm is expected for cuprous oxide \cite{refract}. A wedge shaped sample was fabricated from a synthetic crystal that was grown from 99.99\% Cu \cite{chang2013removal}. The wedge was designed to examine coherence lengths \cite{boyd1971linear,meredith1981cascading} between 6 and 300 \begin{math}\mu\end{math}m. However, a lack of beating (from destructive interference within the third harmonic) showed that the coherence length is not in the range we examined.

\begin{figure}
	\begin{center}
		\includegraphics[scale=.69]{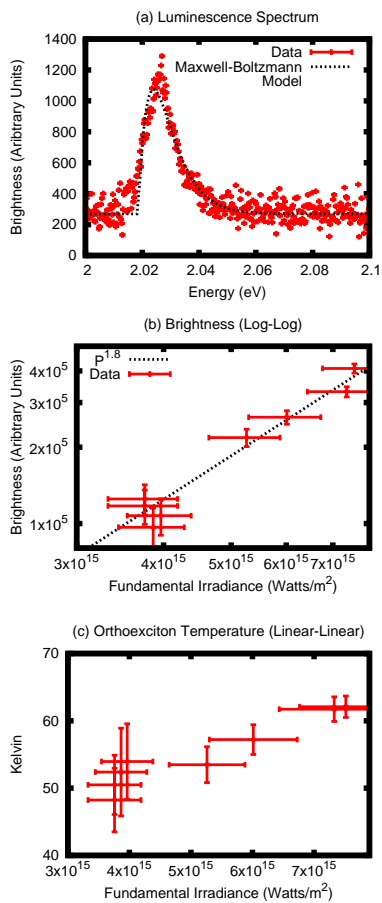}
	\end{center}
	\caption{(a) Luminescence spectrum from three photon absorption of 1400 nm light, using the highest laser power.  (b) Log-Log representation of the dependence of the luminescence intensity as determined from a Maxwell-Boltzmann model on various laser peak irradiances. (c) Linear representation of the dependence of the temperature of the exciton gas on various laser peak irradiances.}
	\label{fig:lum}
\end{figure}

	The wedge-shaped sample was cooled to 2.7 Kelvin and the luminescence spectrum due to third harmonic absorption was measured in the transmission geometry using a 1,400 nm pump (Figure \ref{fig:lum} (a)) where the sample was 0.3 mm thick.  In this pumping and detection configuration, the primary luminescence is due to decay of an orthoexciton into a photon and a $\Gamma_{12}^-$ phonon.  The luminescence does not form a beam.  The spectrum yields the time-averaged thermal distribution of excitons.  The time averaged effective exciton temperature, which is distinct from the lattice temperature, was determined by comparing the Maxwell-Boltzmann distribution to the spectrum.  
	The dependence of the intensity of the luminescence, as determined from the model, on the fundamental laser peak irradiance follows a power law with exponent $1.8\pm 0.2$ (Figure \ref{fig:lum} (b)), consistent with the z-scan results \cite{mani2009large}.
	{Exciton decay into luminescence is not necessarily linear in exciton density \cite{mani2010nonlinear,frazer2013unexpectedly}.}
	The effective exciton temperature (Figure \ref{fig:lum} (c)) should be higher under more intense excitation.  

 A scheme  has been developed for reducing three photon processes when only two-photon absorption is desired which involves \begin{math}
	 \pi
 \end{math} phase shifting half the pump photons\cite{ideguchi2008coherent}.  No specific claims are made about the size of the reduction.  Assuming third harmonic generation dominates over photoionization, Result (\ref{efficiency}) shows that 40\% of the third harmonic generation remains.

Cuprous oxide is a promising material for applications that require optical third harmonic generation while simultaneously having a low second harmonic background.  To our knowledge, no second harmonic generation has been observed in cuprous oxide crystals, despite extensive studies of two-photon absorption.

Third harmonic generation can also be used as a diagnostic tool for cuprous oxide.  The study of stress traps for excitons in cuprous oxide has been a popular field of research \cite{trauernicht1986thermodynamics,yoshioka2012relaxation,schwartz2012dynamics}.  The potentials of these traps can be mapped out by utilizing the localized nature of tomographic third harmonic generation absorption microscopy.  Inclusions \cite{schmidt1974growth} commonly induce stress in the crystal, which can also be measured by determining the resulting localized shift in exciton energy levels.

In conclusion we note that most third harmonic studies involve cascaded \begin{math} \chi^{(2)} \end{math} processes \cite{eimerl1997multicrystal}.  Cuprous oxide is an interesting example of a purely \begin{math} \chi^{(3)} \end{math} system.  Previously, at low temperature the transparent region of cuprous oxide was inaccessible to third harmonic studies.  Strong signals have now been observed in that region.  Here third harmonic generation in cuprous oxide has been explored, including absolute intensity measurements.  The anisotropy of the harmonic generation in this cuprite structured crystal has been shown to follow the cubic crystal pattern.  The results will be useful for exciton studies and, possibly, applications based on third harmonic generation or, more generally, \begin{math} \chi^{(3)} \end{math}.

We gratefully acknowledge NSF IGERT DGE-0801685.  Use of the Center for Nanoscale Materials was supported by the U. S. Department of Energy, Office of Science, Office of Basic Energy Sciences, under Contract No. DE-AC02-06CH11357.  
Crystal growth was supported by NSF DMR-1307698 and in part by Argonne National Laboratory under U.S. Department of Energy contract DE-AC02-06CH11357.
This work made use of the X-Ray and OMM Facilities supported by the MRSEC program of the NSF (DMR-1121262) at the MRC of Northwestern.

\clearpage

\section*{Informational Fifth Page}

\end{document}